\documentclass{aa}  

\usepackage{graphicx}
\usepackage{txfonts}

\usepackage[usenames, dvipsnames]{xcolor}

\def\ie{\textit{i.e.,}}              

\newcommand{\halpha}{H$\alpha$}
\newcommand{\CaII}{\ion{Ca}{ii}}

\newcommand{\CaIIHK}{\ion{Ca}{ii}\,H\,\&\,K}

\newcommand{\MgII}{\ion{Mg}{ii}}

\newcommand{\SiIV}{\ion{Si}{iv}}
\newcommand{\OIV}{\ion{O}{iv}}
\newcommand{\NiII}{\ion{Ni}{ii}}
\newcommand{\MnI}{\ion{Mn}{i}}
\newcommand{\HeII}{\ion{He}{ii}}
\newcommand{\FeII}{\ion{Fe}{ii}}
\newcommand{\FeIX}{\ion{Fe}{ix}}
\newcommand{\FeXII}{\ion{Fe}{xii}}

\newcommand{\ada}[1]{\color{blue}{#1}}

\begin{document}

   \title{Ellerman bombs and UV bursts: transient events in
     chromospheric current sheets}
   \titlerunning{Ellerman bombs and UV bursts}

   \subtitle{}

   \author{V. Hansteen\inst{1,2}
          \and
          A. Ortiz\inst{1,2}
          \and V. Archontis\inst{3}
          \and M. Carlsson\inst{1,2}
          \and T. M. D. Pereira\inst{1,2}
          \and J. P. Bj{\o}rgen\inst{4,1}
          }

   \institute{Rosseland Centre for Solar Physics, University of Oslo,
              P.O. Box 1029 Blindern, NO-0315 Oslo, Norway \\
              \email{viggoh@astro.uio.no}
         \and
             Institute of Theoretical Astrophysics, University of Oslo, P.O. Box 1029 Blindern, NO-0315 Oslo, Norway
         \and
             School of Mathematics and Statistics, St. Andrews University, St. Andrews, KY169SS, UK
        \and
             Institute for Solar Physics, Department of Astronomy, Stockholm University, Albanova University Centre, SE-10691 Stockholm, Sweden
             }

   \date{Received ; accepted}

 
  \abstract
   {Ellerman bombs (EBs), observed in the photospheric wings of the \halpha\ line, and UV bursts, observed in the transition region \SiIV\ line, are both brightenings related to flux emergence regions and specifically to magnetic flux of opposite polarity that meet in the photosphere. These two reconnection-related phenomena, nominally formed far apart, occasionally occur in the same location and at the same time, thus challenging our understanding of reconnection and heating of the lower solar atmosphere.}
   {We consider the formation of an active region, including long fibrils and hot and dense coronal plasma. The emergence of a untwisted magnetic flux sheet, injected $2.5$~Mm below the photosphere, is studied as it pierces the photosphere and interacts with the preexisting
   ambient field. Specifically, we aim to study whether EBs and UV bursts are generated as a result of such flux emergence and examine their physical relationship.} 
   {The Bifrost radiative magnetohydrodynamics code was used to model flux emerging into a model atmosphere that contained a fairly strong ambient field, constraining the emerging field to a limited volume wherein multiple reconnection events occur as the field breaks through the photosphere and expands into the outer atmosphere. Synthetic spectra of the different reconnection events were computed using the $1.5$D RH code and the fully 3D MULTI3D code.}
   {The formation of UV bursts and EBs at intensities and with line profiles that are highly reminiscent of observed spectra are understood to be a result of the reconnection of emerging flux with itself in a long-lasting current sheet that extends over several scale heights through the chromosphere. Synthetic spectra in the \halpha\ and \SiIV\ 139.376~nm lines both show characteristics that are typical of the observations. These synthetic diagnostics suggest that there are no compelling reasons to assume that UV bursts occur in the photosphere. Instead, EBs and UV bursts are occasionally formed at 
   opposite ends of a long current sheet that resides in an extended bubble of cool gas.}
   {}


   \maketitle
%

\section{Introduction}

Emerging flux regions are host to a large variety of transient phenomena that are driven by the 
interaction of the emerging field with the photospheric, chromospheric, and 
coronal plasma, the preexisting ambient field, and of the field with itself. Such interactions 
are a necessary part of the rise of the field into the corona and the forming magnetic field of the active region; the rising field carries with it considerable mass, which must fall back to and through the photosphere in order to allow the field to attain coronal heights. 
Reconnection plays an 
important role in this process, cutting field lines so that dense material can 
fall while at the same time alleviating the weight on the field lines and allowing them to form the longer loops that make up the active region corona and chromosphere of the active region \citep[see,  e.g.,][]{2004ApJ...614.1099P}. 

These dynamical phenomena include Ellerman bombs 
\citep[EB hereafter,][]{1917ApJ....46..298E} and UV 
bursts \citep{2014Sci...346C.315P}, which both occur in the 
vicinity of merging or cancelling photospheric fields of opposite 
polarities. Ellerman bombs are first and foremost recognized by 
strong emission in the wings of \halpha, with little or no visible 
signal in the line core. This indicates that EBs are formed at 
photospheric or upper photospheric heights, at most some few hundred kilometers 
above the photosphere. As observed in spectroheliograms in the wings 
of \halpha,\ they appear as flame-like structures that jut out of 
the photosphere when seen toward the limb \citep{2011ApJ...736...71W}. 
The primary signature of UV bursts is found in small ($\sim1\arcsec$) extreme brightenings in 
the \SiIV\ lines that have been observed with the Interface Region Imaging Spectrograph (IRIS) \citep{2014SoPh..289.2733D} to be as much 
as two to three orders of magnitude brighter than the average emission. 
This brightening is accompanied by large line broadening, of order 
$200$~km/s or more, and importantly, superimposed absorption features by lines associated with 
much cooler temperatures such as from 
\NiII\ and \FeII\ lines. UV bursts and EBs are also found at sunspot light bridges \citep[e.g.,][]{2015ApJ...811..137T} and in moving magnetic features \citep[e.g.,][]{2015ApJ...809...82G}.

The observed relationship between EBs and UV bursts has recently been 
the focus of debate and several papers,
see \citet{2015ApJ...812...11V},
\citet{2015ApJ...810...38K},
\citet{2016A&A...593A..32G},\citet{2016ApJ...824...96T}, and \citet{2018SSRv..214..120Y},
and references cited therein, and most recently, \citet{Ortiz_etal2018}. 
The question is whether EBs and UV bursts are connected, and if so, how they are related. The 
height at which the plasma temperature becomes high enough to emit radiation in the 
\SiIV\ lines also needs to be determined. \citet{2014Sci...346C.315P} 
have speculated on the possibility of a relationship between UV bursts and 
EBs. These authors were unable to conclude whether UV bursts and 
EBs were the same phenomena or not. \cite{2015ApJ...808..116J} suggested that the \SiIV\ emission was more naturally explained as stemming from the chromosphere and a result of Alfv{\'e}nic turbulence.
\citet{2015ApJ...812...11V} noted that all \halpha\ fitting exercises 
show that EBs represent temperature enhancements of the low
standard model chromosphere by at most a few thousand  Kelvin ot
less. In addition, they claim that \SiIV\ intensities with the observed
properties cannot be obtained from any of these models.  On the other hand, 
based on \OIV\ emission, \MnI\ absorption in the
wings of \MgII, enhanced emission in the \MgII\ wings but not core, deep 
absorption in \NiII\ lines, and compact brightenings in the AIA~170~nm passband, 
\citet{2016ApJ...824...96T} concluded that the UV bursts that occur in 
connection with EBs are formed in the photosphere. While further
noting that other UV bursts may be formed in the chromosphere.
\citet{2017RAA....17...31F} point out that the observed \halpha\
emission cannot be reproduced using non-local thermal equilibrium
(non-LTE)
semi-empirical modeling if the temperature is above $10\,000$~K.
They suggested that the coincidence of EBs and UV bursts could be due a 
projection effect. Furthermore, 
\ \citet{Ortiz_etal2018}, for example, found that EBs and UV bursts can be cospatial 
and cotemporal, but can also occur independently of each other in both 
time and location. The occurrence of cospatial and cotemporal EBs and UV 
bursts in the same spectral image is between 10\%\ and 20\% of all observed 
\halpha\ brightenings in that image. When these two reconnection events 
do occur together, they may do so nearly simultaneously or with a delay 
of a few minutes; the UV burst shows a more rapid rise and fall in intensity
and is more short lived than the cospatial EB.


The question then is whether and how models can naturally reproduce 
the required conditions to generate EBs and UV bursts in the 
same location and at the same time. Modeling reproduces the diagnostic 
signatures of EBs and UV bursts as a result of large angle 
reconnection, essentially of  fields oriented in diametrically 
opposite directions. EBs result when the reconnection occurs in the 
upper photosphere with temperatures some $1\,000$--$3\,000$~K higher 
than ambient \citep[e.g., ][]{2013JPhCS.440a2007R}, while UV bursts are 
reproduced when the reconnection is 
located higher in the chromosphere and entails much higher temperatures 
than those found in EB models, at least high enough to ionize silicon 
three times: 20~kK if in dense photospheric material, 80~kK at 
lower densities, as pointed out by \cite{2016A&A...590A.124R}.
It may be possible to generate temperatures as high as $25$~kK through
reconnection in the upper photosphere or lower chromosphere, as shown
by
\citet{2016ApJ...832..195N,2018ApJ...852...95N}. 

\citet{2018ApJ...862L..24P} and \citet{2019ApJ...872...32S} pointed out in a theoretical work that flux cancellation can cause simultaneous reconnection at different heights in the solar atmosphere. \citet{2017ApJ...839...22H} found that both EBs and UV bursts were generated in a model where a flux sheet was injected some 2.5~Mm below the photosphere and allowed to emerge through the photosphere into an essentially magnetic field 
free corona. Upon reaching the photosphere, no longer buoyant, the 
field stalled, but eventually emerged in locations of strong field, 
filling the chromosphere with expanding cool bubbles. These bubbles 
pushed the original weak-field outer atmosphere aside, filling a large 
area in height with cool gas. As the bubbles grew
horizontally, they came into contact with each other, typically 
with oppositely oriented fields, and formed current sheets. These 
locations were the sites of EBs and UV bursts. 

   \begin{figure*}
   \centering
   \includegraphics[width=\hsize]{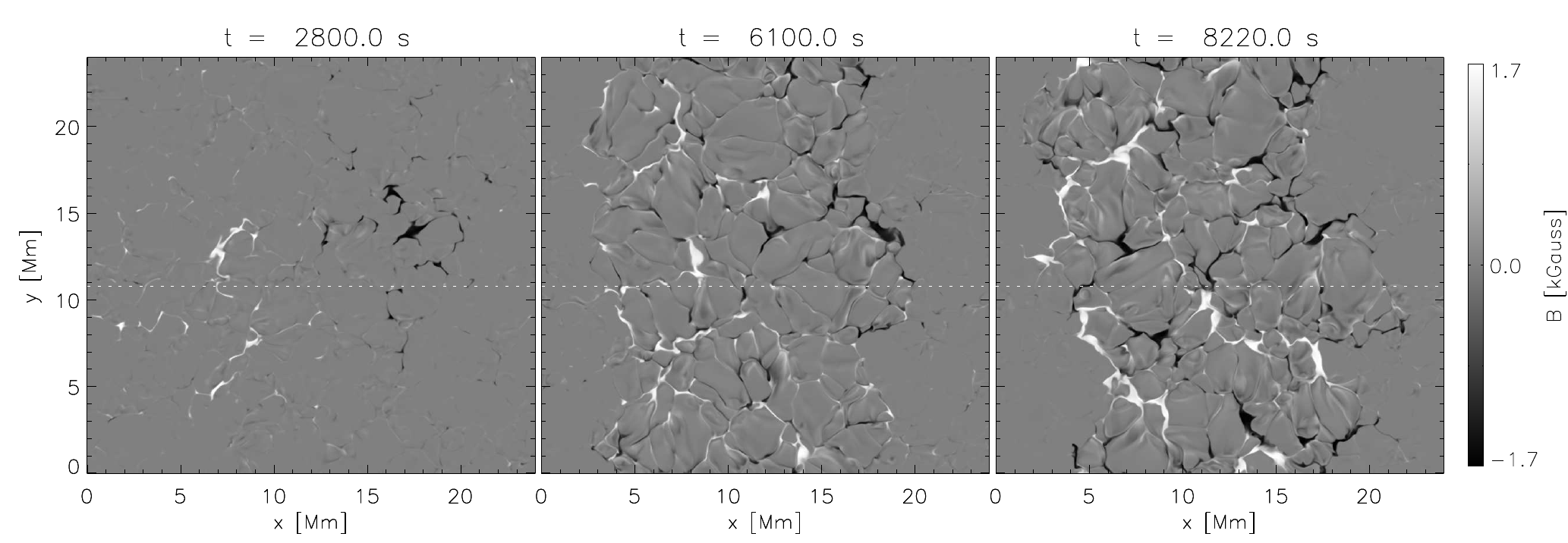}
   \includegraphics[width=\hsize]{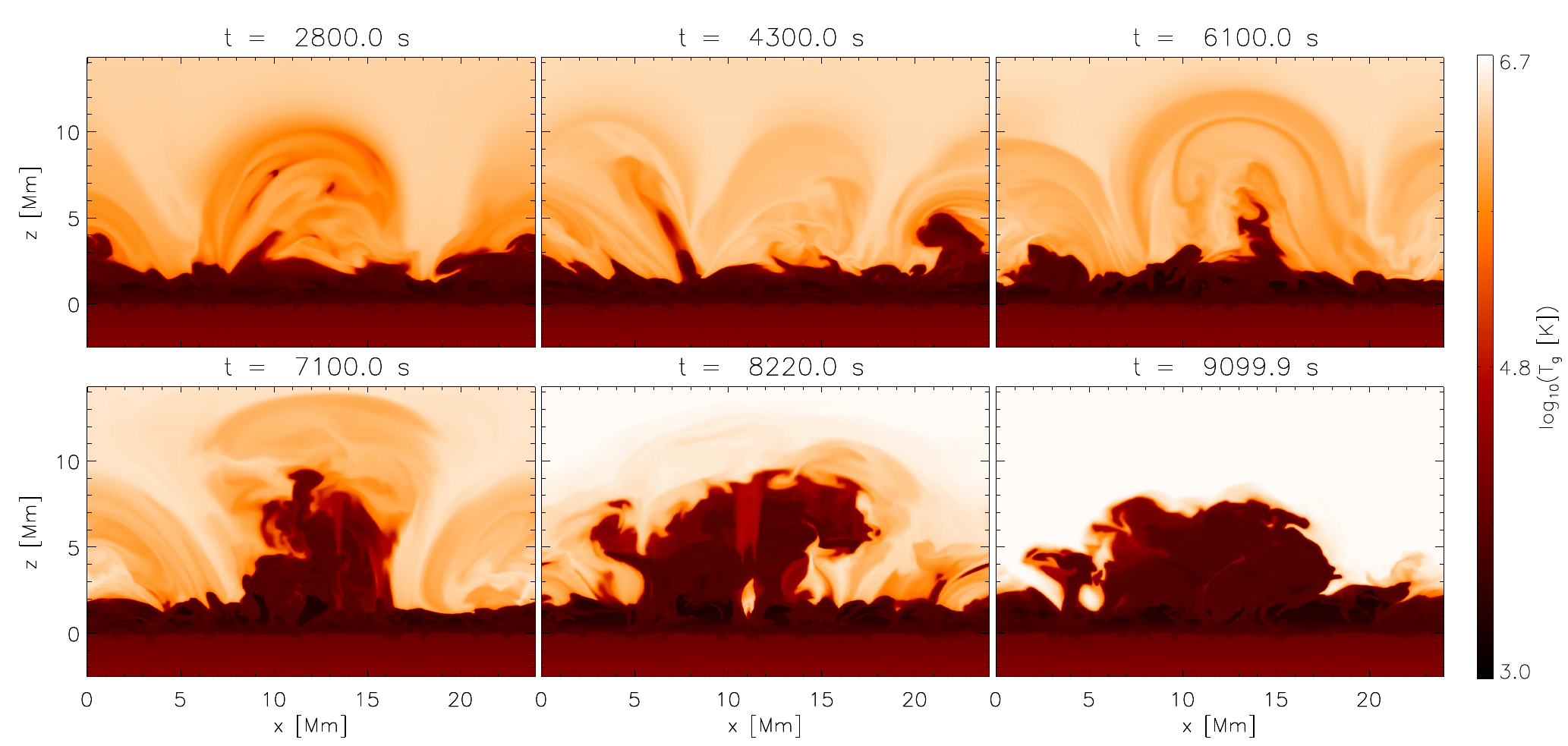}
   \caption{Vertical magnetic field $B_z$ in the photosphere at three
     separate times during the simulation (upper panels). The white
     dotted lines at $y=10.78$~Mm in the upper three panels show the
     location of the vertical cut in the $xz$-plane of the
     chromospheric and coronal temperature structure displayed in the
     panels below. These six panels show the evolution of the outer
     atmosphere as the magnetic bubble pierces the photosphere and
     expands through the chromosphere, pushing the corona upward,
     carrying cool material to large heights and heating the corona as
     the emergent and ambient field interact at large angle. The EB
     and UV burst described in the paper both occur in a period
     centered on $t=8200$~s. The UV burst visible in the lower middle
     panel lasts some 200~s. The temperature rise of the EB is not
     easily visible in these panels because the temperature of EBs is
     far lower than that of the UV burst (but see
     figure~\ref{fig:ebuv_uztg_EB}). The EB occurs on the white dotted
     line around $x=11.5$~Mm near the center of the meeting opposite
     polarities that are visible in the upper right panel, and it
     lasts significantly longer than the UV burst. It
     starts at $t=7600$ and lasts at least $1200$~s thereafter.}
              \label{fig:ebuv_bz_tg_evol}
    \end{figure*}

However, in that model no colocated or cotemporal EBs and UV bursts were produced.
Rather, it was found that the high velocities and temperatures needed to 
produce UV bursts occurred several chromospheric scale heights above 
the photosphere, while EBs were confined to locations just above the 
photosphere. In this paper a model very similar to that described in 
\citet{2017ApJ...839...22H} is considered, but instead of a weak 0.1 Gauss slanted field, a 
stronger preexisting outer atmospheric field (average field strength 2~Gauss at 10~Mm) 
that is oriented at an angle to the injected emerging field is employed. 
Thus, in addition to improving the numerical 
resolution, the impetus of this work is the expectation that 
the preexisting ambient field helps contain and confine the emerging field. This
should give rise to large angle reconnection and thus powerful coronal 
heating as the fields eventually interact, forming a typical realistic
active region chromosphere covered by a hot corona. We do indeed 
find the formation of long chromospheric fibrils in this model, and
significant
coronal heating. The average temperature at 10~Mm initially is 2~MK, but rises in the span
of 30 minutes to almost 10~MK before falling again over the next hour to the 4--5~MK that  are typical of 
active regions \citep{2012ApJ...759..141W}. In addition, we also find that both EBs and UV bursts are produced.
Occasionally and critically, we find that these phenomena are found to be both colocated and cotemporal, as observed, for instance, by
\citet{2015ApJ...812...11V}, \citet{2016ApJ...824...96T},
 and \citet{Ortiz_etal2018}. Our simulations 
seem to indicate that the concurrence of EBs and UV bursts are due to a long 
vertical or nearly vertical current sheet with intensive reconnection that stretches 
from the upper photosphere, where EB signatures are generated, to several thousand kilometers upward. UV bursts are the result of reconnection across several chromospheric scale 
heights well above the photosphere. Line and continuum absorption is provided by cool
gas that is carried into a vastly expanded chromosphere as the emerging field breaks 
through the photosphere and rises to fill the nascent active region corona.


   \begin{figure}
   \centering
   \includegraphics[width=7cm]{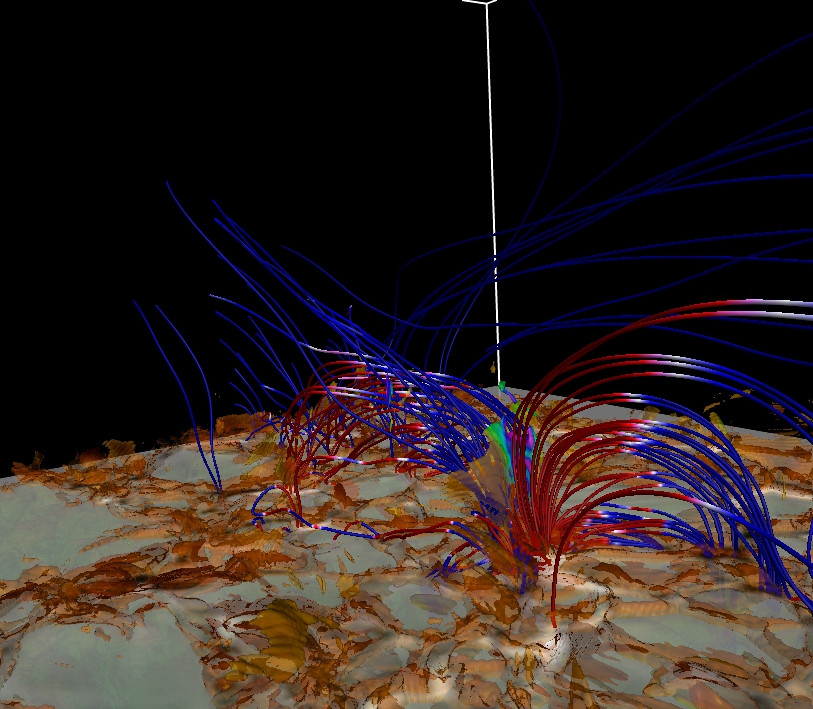}
   \includegraphics[width=7cm]{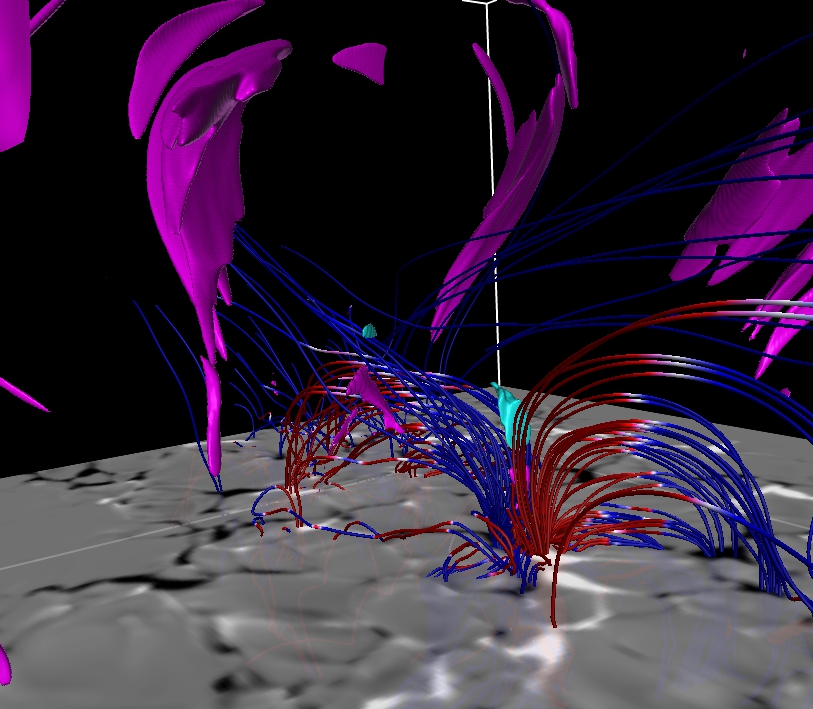}
   \includegraphics[width=7cm]{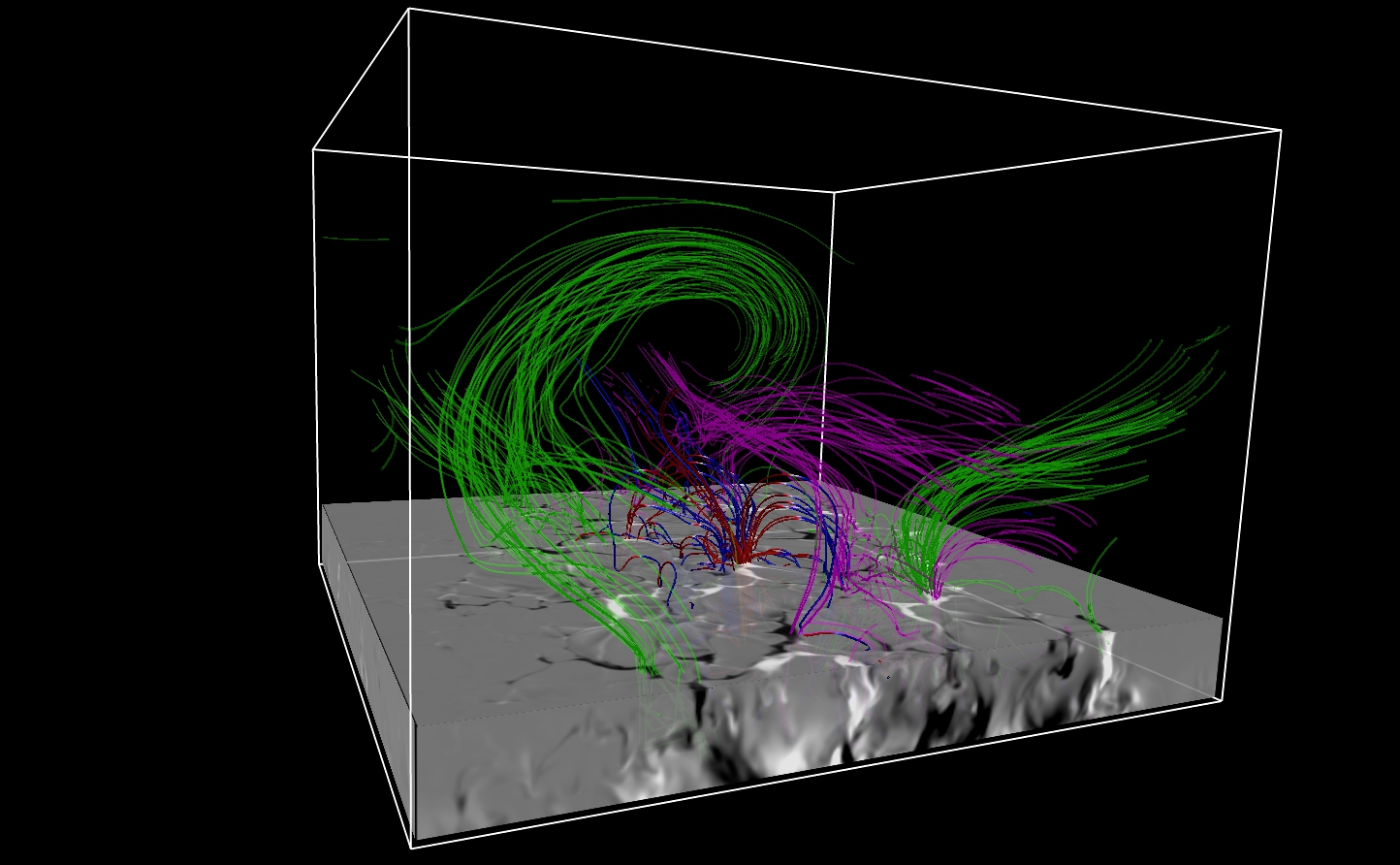}
      \caption{Structure of chromosphere and corona above the flux cancellation site. Joule heating (top panel), up- and downflows in and around the current sheet (center panel), and magnetic field lines surrounding and above the current sheet.  The green field lines in the lower panel outline the original ambient field that has been pushed upward and compressed by the emerging field (purple field lines) that is oriented largely in the $y$-direction  (front to back in the figure). 
               Below this field, we find a newly emerged field, drawn in red and blue to mark positive and negative $B_z$. The reconnection site driving the EB and UV burst is evident in the center of each panel. Reconnection heats and accelerates the plasma. The center panel shows isosurfaces of $100$~km/s (cyan) upward flows and $-60$~km/s downward (purple) flows. Beige, brown, and green in the upper panel show the intensity of the Joule heating.
              }
         \label{fig:ebuv_3D}
   \end{figure}

   \begin{figure*}
   \centering
   \includegraphics[width=\hsize]{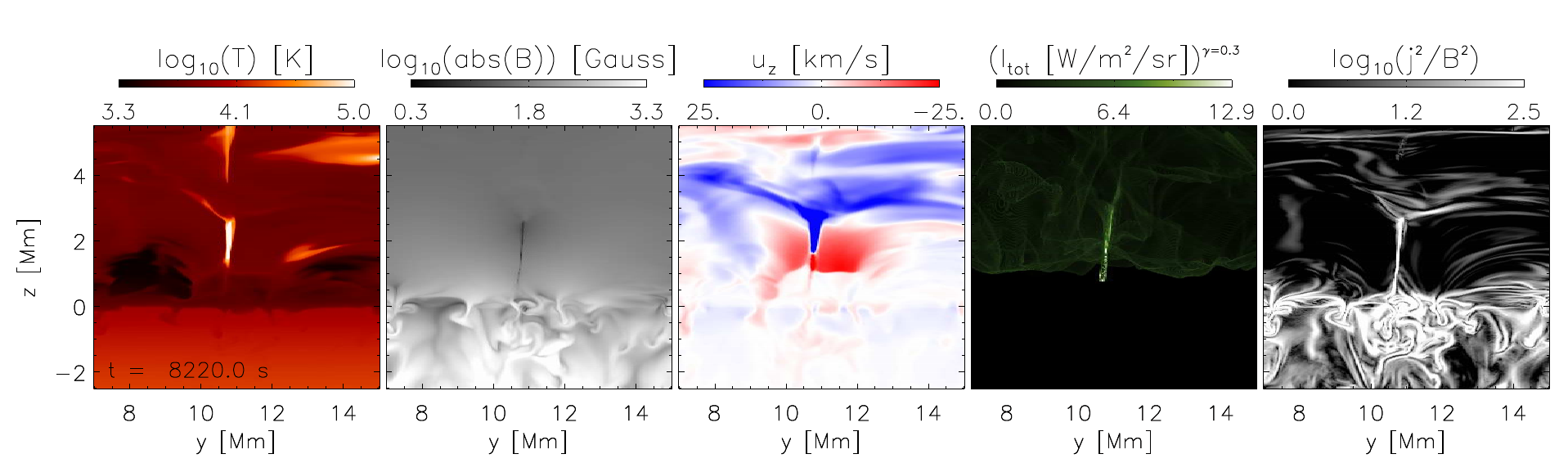}
      \caption{Vertical cuts, at $x=10.78$~Mm in the $yz$ plane, of regions in the vicinity of the current sheet of figure~\ref{fig:ebuv_bz_tg_evol} at 
       $t=8220$~s near the time of maximum intensity of the \SiIV\ lines. {\ada From left to right:} log temperature, log total magnetic field strength, vertical velocity, 
               \SiIV\ 139.376~nm emission, and current density (${j^2/B^2}$). The temperature is saturated at $\log(T)=5$ and the velocity 
               at $u_z=[-25,25]$~km/s, the maximum values of these quantities are higher than plotted here, as shown in figure~\ref{fig:ebuv_uztg_UV}.
              }
         \label{fig:emergence_side}
   \end{figure*}

\section{Methods}

\subsection{Initial model}

The initial model we used for this study is a version of the publicly available
Bifrost model \citep{2016A&A...585A...4C} with higher resolution that covers the same spatial extent and has roughly  
the same magnetic topology. This new model was run 
for roughly one hour solar time and is designed to have an effective 
temperature of $T_{\bf eff}\approx 5780$~K, convective energy transport 
below the photosphere, and radiative losses maintained by the injection of 
entropy at the bottom boundary. The outer atmosphere is maintained by 
acoustic shocks generated from motions in the convection zone and by the 
interaction of convective motions and the magnetic field. 

The average field strength in the photosphere is 50~Gauss originally, and 
at the onset of the run, the 
vertical component of the photospheric field is concentrated in two 
amorphous regions, positive polarities between $x=[6,10]$~Mm 
and $y=[5,15]$~Mm, and negative polarities some 10~Mm distant 
between $x=[15,19]$~Mm and $y=[12,17]$~Mm, 
as shown in the left panel of Figure~\ref{fig:ebuv_bz_tg_evol}. 
Both polarities have maximum strengths of about $1\,500$~Gauss in the 
photosphere, and chromospheric and coronal field lines stretch between them.
Hints of these field lines are visible in the temperature structure of 
the lower corona shown in the middle left panel of  
Figure~\ref{fig:ebuv_bz_tg_evol}. Coronal field lines, for example, as 
outlined in emission in the \FeXII\,19.51~nm line formed at $1.5$~MK,
show that the field lies at a $15^{\circ}$ angle to the $x$-axis.

The photosphere, as defined by the height where 
$\tau_{500 {\rm nm}}=1$, is located at $z=0$~km.
Above it, though corrugated, the initial chromosphere extends to
2~Mm, at which height we find the rapid temperature rise to coronal 
temperatures. Before the newly injected magnetic flux breaks 
through the photosphere, we find average coronal temperatures of 2~MK 
at 10~Mm. Chromospheric and coronal temperatures and dynamics are 
maintained by acoustic waves and by footpoint braiding and subsequent
nanoflares resulting from convective motions, as described in 
\citet{2016A&A...585A...4C} and \citet{2015ApJ...811..106H}.  

The model extends from $2.5$~Mm below the photosphere to $14.5$~Mm above 
the photosphere in the vertical direction, and $24$~Mm in both horizontal 
directions. This computational domain is covered by 
$768\times768\times768$ grid points, giving a horizontal grid size of $31.25$~km. The grid size varies
in the vertical direction and is roughly $12$~km from 
$-1.0$~Mm below the photosphere to $4.5$~Mm through the chromosphere. It slowly 
increases where the scale heights are larger, to $80$~km at the upper boundary 
in the corona and to $20$~km at the bottom convection zone boundary. 

The calculations were carried out using the Bifrost 
code \citep{2011A&A...531A.154G}. This code solves the magnetohydrodynamics (MHD) equations for a 
plasma in which the ionization state, for this specific model, is assumed to be in LTE.  Energy 
source and sink terms include radiative losses; solving the equations of radiative transfer in 
four frequency bins and scattering for the optically thick photosphere and lower 
chromosphere; optically thin radiative losses based on recipes derived by \citet{2012A&A...539A..39C} 
for the upper chromosphere, transition region, and corona;
and 
thermal conduction along the magnetic field. The recipe for optically thin radiative losses 
includes back-heating by the Ly$\alpha$ line and the Lyman-continuum, as described 
in \citet{2012A&A...539A..39C}. 


\subsection{Spectral synthesis}
In order to compare our simulation with observed EBs and UV bursts, we synthesized spectral data in several often observed lines that are formed in the photosphere, chromosphere, transition region, and corona. These lines include \halpha, the \CaIIHK\ and \CaII\ 854.2~nm triplet line, lines of \SiIV\ including the 139.376~nm line, and the coronal \FeXII\ 19.51~nm line.

The \halpha\ scattering line requires a full 3D evaluation of the radiation field to obtain chromospheric features in the emergent intensities \citep{2012ApJ...749..136L}.
We therefore calculated the hydrogen lines using the MULTI3D code \citep{2009ASPC..415...87L}. To model H$\alpha,$ we used a three-level plus continuum model atom of H \textsc{i}, where Ly$\alpha$ is treated with a Doppler absorption profile in complete redistribution \citep{2018_bjorgen}. 


Synthetic \CaII\ line profiles were calculated using the RH 1.5D code 
\citep{2015A&A...574A...3P}. RH 1.5D 
solves the non-LTE radiative transfer problem by treating each column 
of a 3D model as a 1D atmosphere (1.5D approximation). 
Based on the RH code \citep{2001ApJ...557..389U}, RH 1.5D allows partial 
redistribution effects, multiple species treated 
in non-LTE, and adding blending lines. To synthesize \CaII,\ we used a five-level plus continuum model atom in which the \CaIIHK\ lines were 
treated in partial redistribution, while the infrared triplet lines were 
treated in complete redistribution.

Although the \SiIV~139.376~nm line, formed at transition region temperatures (80~kK), is usually 
optically thin, the 
opacity of the line can sometimes become higher than unity. When the line was viewed from directly above, we therefore computed the \SiIV\ lines by solving the non-LTE equations of radiative transfer using the RH 1.5D code. 
We employed a 9-level atom, including 5 levels of \ion{Si}{iii}, the \ion{Si}{iv} ground state, the upper levels of the
139.376 and 140.277~nm lines, and the \ion{Si}{v} ground state. In addition, we included two more active atoms: a model atom with 
15 levels of \ion{Si}{i} and the ground state of \ion{Si}{ii,} and a 4-level model of \ion{Ni}{ii}-\ion{Ni}{iii}. 
This allowed us to derive the correct continuum intensity level for the \ion{Si}{iv} lines (the background opacity there is 
dominated by photoionization of \ion{Si}{i}) and 
possible absorption in the \NiII\ 139.332~nm line, which is located at 93~km/s in the blue 
wing of \SiIV~139.376~nm. The \NiII\ line is formed at $14$~kK, which is much lower than the expected formation temperature of \SiIV. 
We did not include the \FeII\ lines that formed in the vicinity of 139.376~nm in this study.
Moreover, the \SiIV\ lines, both as viewed from above and from the side, were also calculated using an optically thin approximation with the same method as described below for \FeXII, and 
with data from version 8 of the CHIANTI atomic database \citep{2015A&A...582A..56D}.
Although differences are visible between the optically thin and optically thick results, the total intensities are 
qualitatively very similar. 

The \FeXII~19.5~nm that forms around 1.5~MK can be considered optically thin, and we calculated this line using 
collisional and ionization state data from version 8 of the CHIANTI database. The contribution to the intensity at frequency 
$\nu$ is given by 
\begin{equation}
    dI_\nu=\phi_\nu(u_z,T)A_{\rm Fe}n^2_{\rm e}g(T)e^{-\tau}dz
,\end{equation}
where $\phi_\nu$ is the (Gaussian) emission profile, $u_z$ is the vertical velocity, $T$ is the temperature, $A_{\rm Fe}$ is the 
abundance of iron, $n_{\rm e}$ is the electron density, $g(T)$ gives the excitation and ionization state of \FeXII, 
and $dz$ is a distance element along the (vertical) line of sight.
Because several regions of potential \FeXII\ emission are separated in places by several megameter of cool gas, we 
included absorption by neutral hydrogen and helium and singly-ionized helium in the $e^{-\tau}$ absorption factor, 
following the recipes of \citet{2005ApJ...622..714A}, as done earlier and described in greater detail in 
\citet{2009ApJ...702.1016D}. The total intensity in the line was found by integrating along the line of sight and in 
frequency across the line profile. 

We assumed ionization equilibrium when we computed the \SiIV\ and \FeXII\ lines. We included the effects of three-body recombination, which would allow \SiIV\ to be formed at temperatures down to 10--20~kK at photospheric densities, but we did not see such temperatures near the dense plasma that forms the photosphere in this simulation \citep[see, e.g., ][]{2016A&A...590A.124R}.



\section{Results}

\subsection{Injection of the magnetic flux sheet}
A magnetic sheet, spanning $x=[4,18]$~Mm and the full range along the $y$-axis,
was injected at the bottom 
boundary 2.5~Mm below the photosphere with a field strength of 
$B_y=2\,000$~Gauss and oriented in the $y$-direction. Through the action 
of buoyancy and convective upflows, the field rose through the upper
convection zone and reached the photosphere after roughly one hour 
(at $t\approx3\,500$~s simulation time). 
On rising, the sheet was perturbed by convective motions; it was pulled down in locations of 
descending plumes of cold plasma and rose more slowly in upflow regions. It therefore reached the surface at slightly different times in discrete locations. 
On arriving at photospheric heights, the field was no longer buoyant 
and the upward motions stalled, but eventually, in locations 
where the field was strong enough, the field broke through the surface. 
The field in these locations rose rapidly and expanded into the chromosphere, carrying cool 
magnetized photospheric material upward. It
was anchored to the photosphere 
in downdrafts where the field became nearly vertical 
\citep[see][for a detailed description of this phase]{2014ApJ...781..126O}.
The state of the vertical field component in the photosphere is 
shown for two separate times in the middle and right panels 
of Figure~\ref{fig:ebuv_bz_tg_evol}. Patches of opposite polarity are seen 
to pierce the photosphere in a fairly complex granular pattern 
that approximately fills the same horizontal extent as the horizontal size 
of the injected flux sheet. The field orientation is also close 
to what it was when it was injected $2.5$~Mm below. It forms an angle of approximately $75^{\circ}$ 
 with the preexisting ambient field, which ensures large-angle reconnection
as these fields, old and new, interact. 

The field that rose into the chromosphere was
strong, strong enough to push the preexisting ambient field upward and compress it. 
The newly emerging flux pulled cool photospheric and chromospheric material 
with it to great heights. This rise first halted some 10~Mm above the photosphere, where 
further expansion was slowed by the preexisting field as the emerging and ambient fields attained 
about equal strengths of $30-50$~Gauss at 10~Mm. 
This in contrast with the previous simulation 
of \citet{2014ApJ...788L...2A} and \citet{2017ApJ...839...22H}, where the 
preexisting ambient field was very weak and was completely swept away by the newly 
emerging flux and fully removed the original corona. In the current simulation
the preexisting ambient field was strong enough to hinder the
expansion of the emerging bubbles and the ejection of the initial
corona, which forces the bubbles to interact more strongly as both
horizontal and vertical expansion is suppressed. This results in
stronger and longer lasting current sheets that achieve lengths or
heights of several thousand kilometers.
The evolution of the emerging and expanding magnetic field is evident in the lower panels 
of Figure~\ref{fig:ebuv_bz_tg_evol}, which show that the entire 
horizontal extent of the flux sheet rises and eventually impacts the corona, 
pushing it aside and filling the upper atmosphere with field and cool gas, 
but only up to a height of 10~Mm. The rising field expands from the photosphere in the 
form of cold magnetized bubbles, as described by \cite{2014ApJ...781..126O}. As 
these bubbles expand, they eventually come into contact with each other,
especially in regions where photospheric motions have brought their 
footpoints close together. 

The interaction between the preexisting ambient field and the newly 
emerging field, which occurs near the top of the emerging bubble, essentially 
through high-angle reconnection, causes the
coronal plasma temperatures at heights at and above 
10~Mm to increase 
significantly. These temperatures, at times of about 10~MK as the flux rises 
and of about 4~MK on longer timescales, are much higher than 
those generally achieved by low-angle reconnection that is driven solely 
by photospheric braiding \citep[see, e.g.,  \ ][]{2007ApJ...666..516G,2015ApJ...811..106H}. 
However, we here concentrate on the 
high-angle
reconnection, close to $180^{\circ}$, which occurs at lower heights, 
in the first few thousand kilometers above the photosphere, where oppositely 
directed magnetic fields that largely stem from the newly emerged 
field interact with each other: They heat and accelerate the cold plasma that is carried 
up by the emerging magnetic field.

\subsection{Formation of the current sheet and reconnection}


The oppositely directed magnetic field is brought together as the field expands into the upper chromosphere and by photospheric
motions in many locations in the region that is covered by the emerging flux sheet. 
In particular, we concentrate on the fields in the vicinity of the center 
of the computational domain, located at $x=11$~Mm and $y=11$~Mm, 
which is clearly visible in the center of the upper right panel of 
Figure~\ref{fig:ebuv_bz_tg_evol}. At this location, two footpoints of opposite 
polarity, each the nexus of loop systems that are oriented roughly 
along the $y$-axis that span three to four granules, 
are brought into contact: A current sheet forms starting at $t=6\,000$~s. This 
current sheet strengthens with time, initially slanted but becoming nearly vertical 
in the next $1500$~s, stretching from the photosphere to almost 4~Mm above. 
After formation, the current sheet maintains its shape and remains vertical and 
in the same location for at least the next $1\,000$~s, that is,\ to 
$t=9\,000$~s (or slightly longer). 

The topology of the magnetic field at the height of the UV-burst emission is shown in figure~\ref{fig:ebuv_3D}, which 
also shows how oppositely directed fields are brought together and become the source of strong Joule heating and reconnection 
jets, which accelerate the plasma to high velocities. Above the reconnection region, we find emerged and previously reconnected field lines that have entered the outer solar atmosphere. They are
oriented along the $y$-axis and are aligned with the injected field coming from below.

The current sheet leaves an imprint on the temperature of the rising cool gas and is 
visible as a heated region in the lower central panel of Figure~\ref{fig:ebuv_bz_tg_evol}, 
which shows a $xz$-oriented vertical cut of the atmosphere at $y=10.78$~Mm and
$t=8\,220$~s, as a `leaf-shaped' object. The form of the current 
sheet heating is also visible in the top panel of figure~\ref{fig:ebuv_3D}. 

The general shape and variation in plasma parameters (temperature, magnetic field 
strength, vertical velocity, and $j^2/B^2$) in and around 
the current sheet are further shown in Figure~\ref{fig:emergence_side}, which shows a 
$yz$-oriented cut at $x=10.78$~Mm through the center of the sheet. 
A movie of its evolution, showing the same panels as in 
figure~\ref{fig:emergence_side}, is included in the supplementary material 
to this article. The high-temperature phase of the current sheet, 
with temperatures well above $10$~kK, which is responsible for the UV burst discussed 
below, only lasts for a limited (200~s) time. In contrast, the current sheet itself 
with associated reconnection events last much longer, and the temperature along 
the current sheet is at all times higher than ambient. 

A number of bidirectional jets with velocities of up to $200$~km/s are excited 
along the current sheet as  reconnection commences. These jets heat the plasma, 
thus forming a thin vertical `leaf' of hot accelerated plasma that is located between 
the interacting
bubbles of newly emerged cool gas. The activity engendered by the
reconnection perturbs the plasma along the entire current sheet all the 
way from coronal heights ($3.5$~Mm) and down to the photosphere. The movie of the current sheet evolution also shows that a number of 
plasmoids are generated and are accelerated either upward or downward during 
the lifetime of the current sheet.

\subsection{Ellerman bomb}\label{sec:resultEB}
\begin{figure}
   \centering
   \includegraphics[width=\hsize]{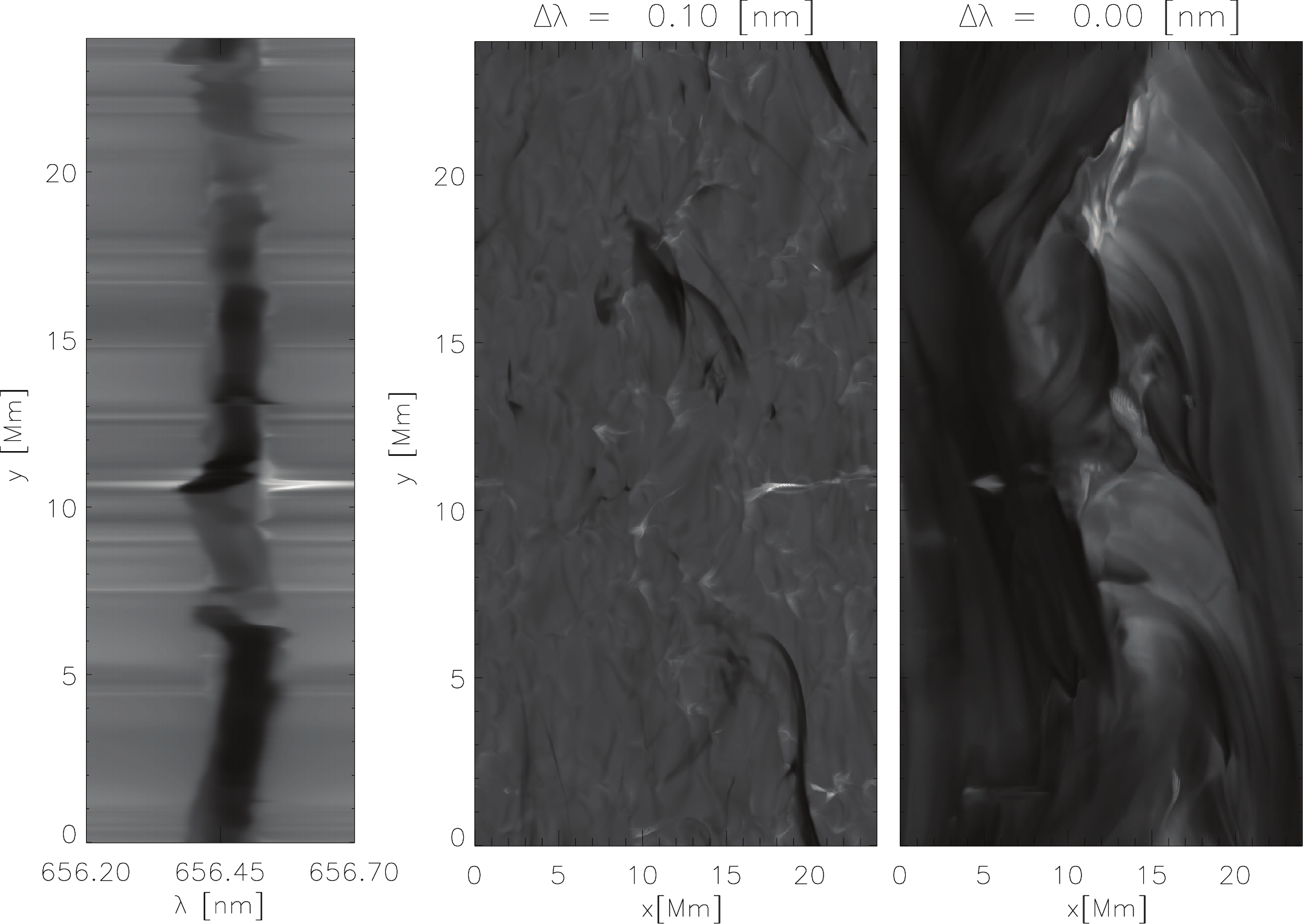}
      \caption{\halpha\ spectroheliograms at $+0.1$~nm and at line center (center and right panels, 
      respectively) at a viewing angle of $\mu=0.5$.  The EB is located at $x=17$~Mm. Because of the projection effect of looking from the side, this corresponds to $x=11$ when viewed from directly above. The line profile over the EB, shown in the left panel, is located at $[x,y]=[17,11]$~Mm and is evident in the line wings.
              }
         \label{fig:ha_profile}
   \end{figure}

In the first few scale heights above the photosphere, this activity leads to the 
formation of an EB. The synthetic \halpha\ spectra calculated using the 
MULTI3D code \citep{2009ASPC..415...87L} display many features that are the same as 
those that are observed. For example, in Figure~\ref{fig:ha_profile} the \halpha\ emission is 
shown as seen from the side at an angle of $\mu=0.5$: the wings of the profile 
form the typical EB moustaches, while the line core shows no brightening 
or evidence of heated gas. The EB forms a flame-like structure that is clearly
visible in spectroheliograms made in the red and blue wings of 
\halpha\ (the red wing is shown in figure~\ref{fig:ha_profile}). This structure
extends to 1.2~Mm above the photosphere. 

The line core, shown in the right panel of figure~\ref{fig:ha_profile},
shows extensive fibrils, some have about the length of the computational 
box, $20$~Mm, but no evidence of brightening directly above 
the site of the EB. The two loop systems that meet and are the source of the EB are 
visible in the center of the right panel of figure~\ref{fig:ha_profile}. They stretch toward 
greater and lower $y$-values from their meeting point close to $y=11$~Mm. 
?

When viewed from directly above, 
the EB region is bright in \halpha  \ in the line core as well, 
but this is due to emission that is formed much higher in the atmosphere, 
at least 5~Mm above the photosphere: 
The cool canopy above the current sheet is heated by 
Ly$\alpha$ and Ly-continuum radiation from the hot plasma of 
the current sheet below, raising the temperature of 
the canopy plasma significantly, as is visible in the lower middle 
panel of figure~\ref{fig:ebuv_bz_tg_evol} and in the left panel of
figure~\ref{fig:emergence_side}. This relatively high-temperature gas 
excites some \halpha\ emission in the region that is heated above the current sheet. 
This heating at large heights directly above the reconnection
region is probably a result of the method we chose for the coronal back-radiation and raises the temperature more than what would occur on the real Sun. The 
heating does not affect the evolution or diagnostics of the EB or UV burst.

\begin{figure*}
   \centering
   \includegraphics[width=0.95\hsize]
   {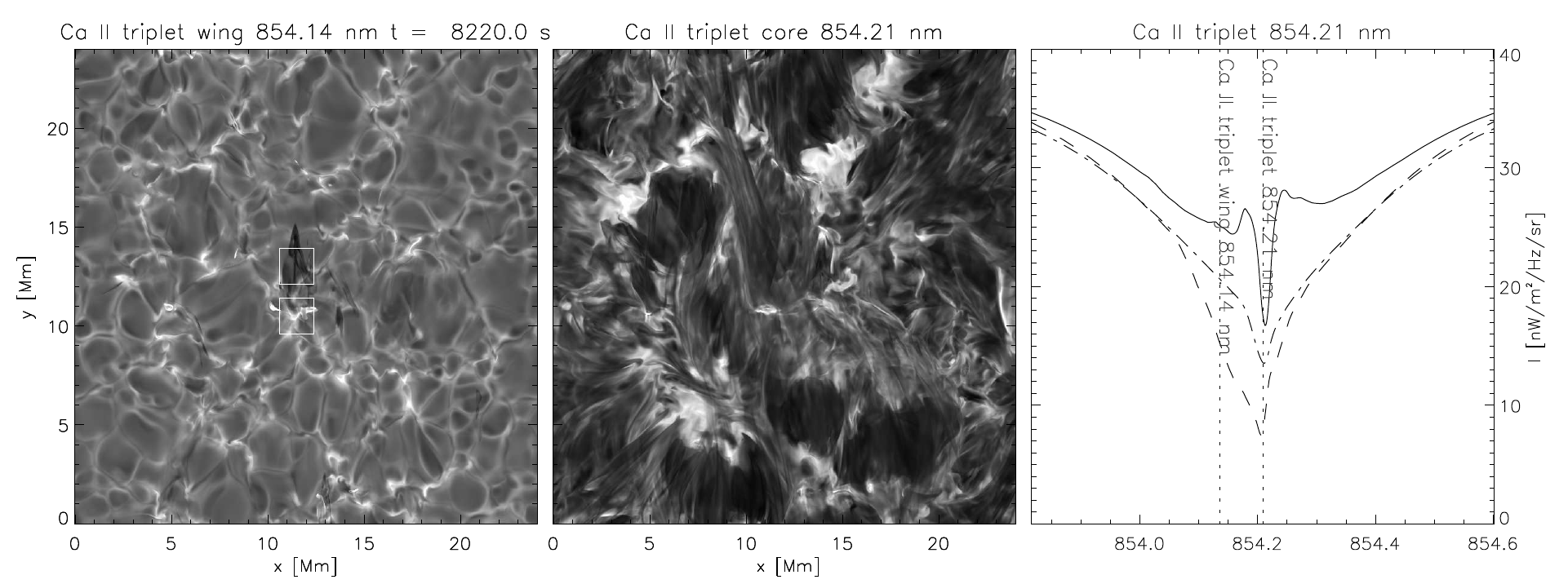}
      \caption{\CaII~854.2~nm line filtergrams showing the blue line wing and line core, as well as the line profile, 
      at the time of a small cool surge. The right panel shows line profiles for the EB region (solid line), 
      the cool surge (dashed line), and the average spectrum (dash-dotted line). The EB and cool surge regions we used 
      to construct the profiles are indicated by the boxes in the left panel.
              }
         \label{fig:ca_triplet_profile}
   \end{figure*}

   Swedish 1-meter Solar Telescope 
   observations \citep{Ortiz_etal2018} show that the wing of the \CaII\ $854.2$~nm line can be a very good proxy for 
the existence of EBs, with enhanced emission that is colocated and roughly cotemporal with bright \halpha\ wing enhancement. 
These observations in the blue wings of both the \CaII\ $854.2$~nm and \halpha\ lines also show dark, 
presumably cool, surges that occasionally emanate from the site of EB emission. 
We find that the EB in this location shows brightening, starting at time $t=7500$~s and 
lasting for at least $1\,200$~s thereafter. This enhanced emission, 0.0735~nm in the 
blue wing of the \CaII\ 854.2~nm triplet line, is shown in 
figure~\ref{fig:ca_triplet_profile}, which also shows emission in the line core and in the 
shape of the line profile. Similar as for \halpha, the cores of the \CaIIHK and the 
\CaII\, 854.2~nm triplet lines show the loops that emanate from the vicinity of the current 
sheet as narrow thin fibrils that stretch toward higher and lower $y$-values, approximately in the $y$-direction. The EB is visible as a narrow band of emission in the line 
wing, and it is also visible at the time of this image ($t=8220$~s) in the line core, although 
this emission is partially covered by overlying longer fibrils. Apparently rising from the 
EB location, about parallel with the positive $y$-axis, a small cool surge is seen 
to be accelerated along the field lines, or loops, that form the cool portion of the canopy. 
Signs of the surge are visible for some minutes (from $t=8140$~s to $t=8300$~s). 
The line profile of the EB region shows enhanced emission in both wings and 
self-reversal in the line core. The profile of the small surge
shows absorption that extends to $0.15$~nm (35 km/s) in the blue wing of 
the \CaII~854.2~nm line. 

\begin{figure}
   \centering
   \includegraphics[width=\hsize]{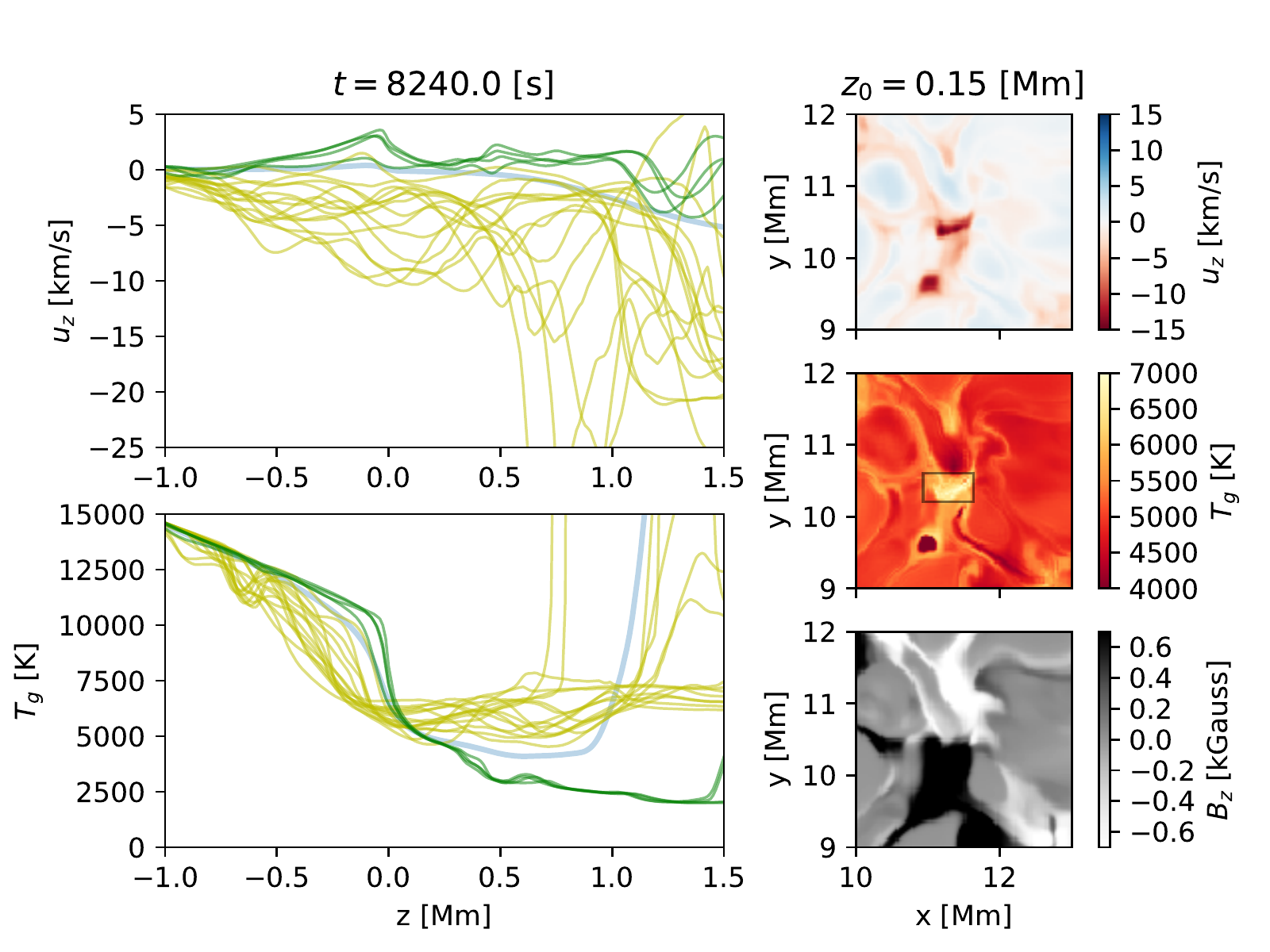}
      \caption{Velocity, temperature, and field strength in the $xy$-plane $z=150$~km 
      above the photosphere (rightmost three panels) in the vicinity of the current sheet 
      that generated the EB described in the text. The current sheet is located in the center of 
      the small box shown in the center of the right central panel. The two leftmost panels 
      show with yellow lines the vertical velocity and temperature as functions of height in 
      vertical columns in this small box, averaged over tiles of 150~km ($5\times 5$ grid zones). 
      The green lines show the average of the same variables some distance away from 
      the current sheet, near $[x,y]=[10,9]$~Mm. The blue lines show the average velocity 
      and temperature as a function of height in the entire computational domain.
              }
         \label{fig:ebuv_uztg_EB}
   \end{figure}

The structure of the atmosphere in the vicinity of the current sheet that produced 
the EB is shown in figure~\ref{fig:ebuv_uztg_EB}. In the location where positive 
and negative polarities meet, that is,{} $x=[11,12]$~Mm, $y\approx 10.5$~Mm, 
we see heightened temperatures. The current 
sheet is the site of high-velocity downflows and hotter-than-average 
chromospheric temperatures that extend all the way down to the photosphere. The 
temperature around the current sheet is about $7\,500$~K, 
at least $1\,500$~K warmer than the ambient average chromosphere. 
(On the other hand, temperatures {\it \textup{below}} the photosphere in 
the location of the current sheet are lower than ambient.)
We find large downflows of 25~km/s or more, which are rapidly 
decelerated in the region down to $500$~km. 
Even closer to the photosphere and below, downflows of about $10$~km/s fill 
the area in and around the current sheet. These higher temperatures and high downflow 
velocities are concentrated in a narrow region with horizontal extent of some 
700~km where field lines of opposite polarity meet, that is,\ in the region of 
the current sheet. In the flux-emerging region outside of the current sheet, 
the chromosphere included in the magnetic bubbles does not generally have 
highly supersonic velocities, nor a temperature rise to coronal temperatures
until the canopy is reached at some 9~Mm above the photosphere.

An exception to this is found in Figure~\ref{fig:ebuv_uztg_EB}, which shows that 
we sometimes also find areas of high downflow velocities, 10~km/s or 
more, in the chromosphere outside the current sheet, such as the circular region near 
$[x,y]=[11,9.5]$~Mm, but without a corresponding 
temperature rise. Temperatures of $<3\,000$~K are seen in regions that
contain large unidirectional fields. These may be the sites of convective 
collapse, with no associated temperature rise and thus no EB.


   \begin{figure*}
   \centering
   \includegraphics[width=0.95\hsize]
   {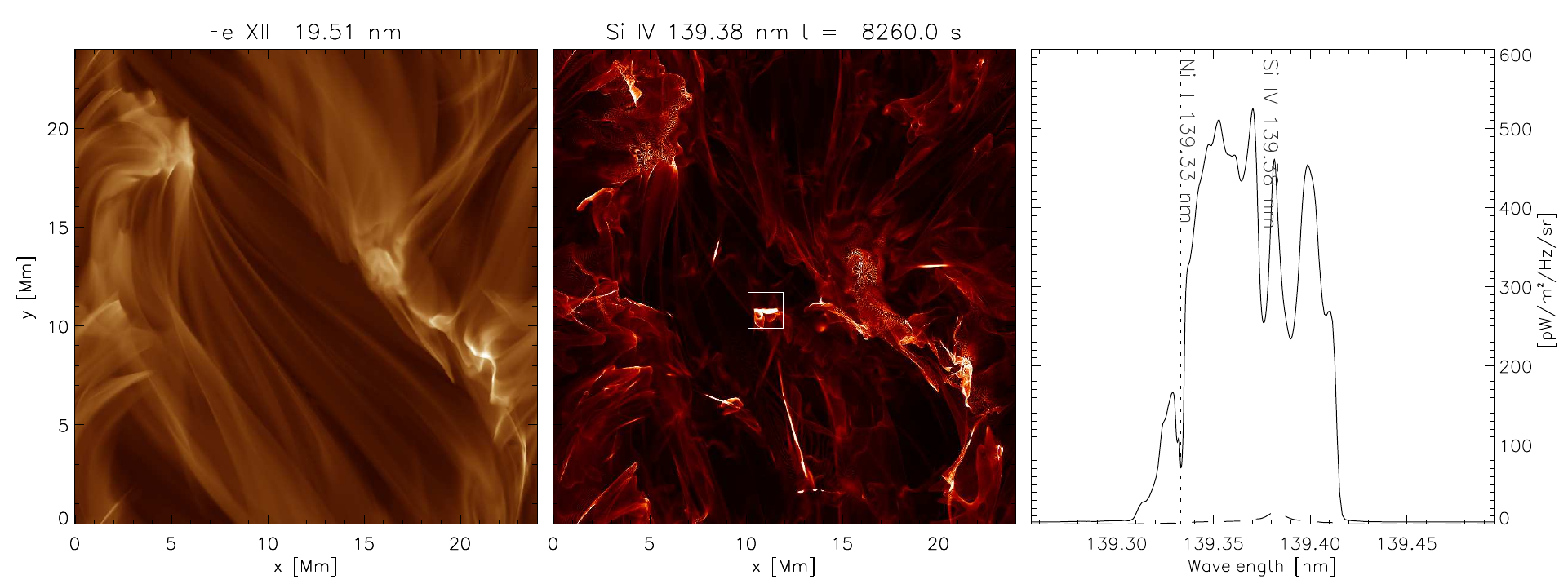}
      \caption{\FeXII~19.5~nm and \SiIV~139.376~nm line total line intensities as well as the line profile 
      of the \SiIV~line (where the full line shows the profile of the UV burst and the dashed 
      line represents the average spectrum). The location of the UV burst is indicated by the box in the central panel.
              }
         \label{fig:si_profile}
   \end{figure*}
   \begin{figure}
   \centering
   \includegraphics[width=\hsize]{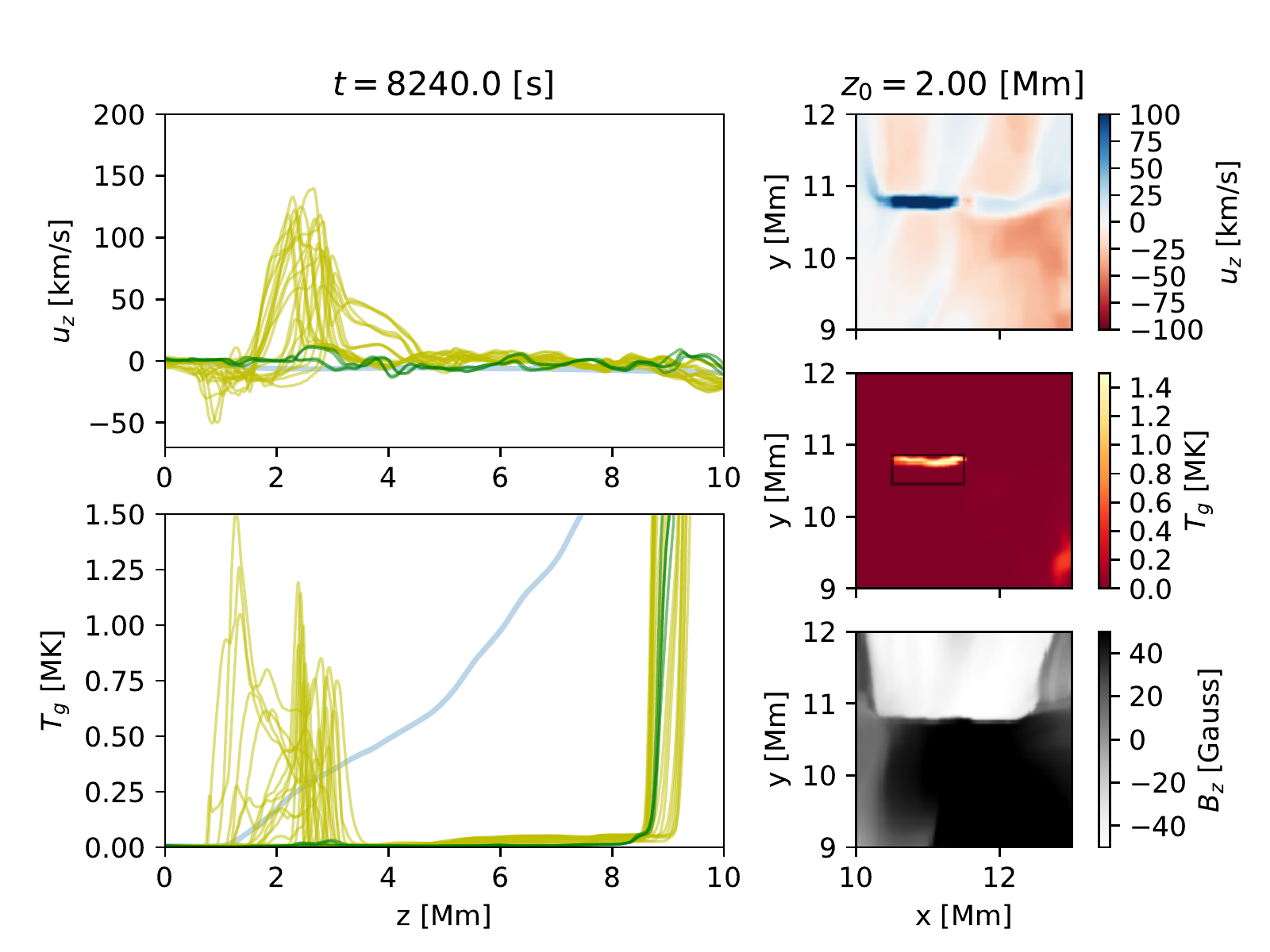}
      \caption{Velocity, temperature, and field strength in the $xy$-plane $z=2$~Mm above the photosphere 
      (rightmost three panels) in the vicinity of the current sheet that generated the UV burst described in the text. 
      The current sheet is located in the center of the small box shown in the center of the right central panel. 
      The two leftmost panels show with yellow lines the vertical velocity and temperature as functions of 
      height in vertical columns in this small box, averaged over tiles of 150~km ($5\times 5$ grid zones). 
      The green lines show the average of the same variables some distance away from the current sheet, 
      near $[x,y]=[10,9]$~Mm. The blue lines show the average velocity and temperature as a function of 
      height in the entire computational domain. 
              }
         \label{fig:ebuv_uztg_UV}
   \end{figure}
\subsection{The UV burst}

While the EB is well established already at $t=7500$~s, 
the UV burst is not visible before $t=8040$~s (although there is an earlier UV 
burst in about the same location that is part of the same flux system) at 
which point the temperature in the current sheet at heights from $700$~km 
up to $3\,500$~km rises rapidly, increasing from $10\,000$~K to more 
than 1~MK within 20~s or less. This event occurs in the same horizontal location
and as part of the same magnetic flux system that produces the EB described in the 
previous section. As described below, this reconnection event leads to a increase in the 
\SiIV\ emission of more than two orders of magnitude above average total intensities. 
Above the current sheet, the plasma remains cool, $<10\,000$~K all the way 
up to 9~Mm above the photosphere, where it rapidly rises to coronal values. 

The plasma is at times heated to above 1~MK, and in principle, emission might also be expected in the \FeXII\ $19.51$~nm line from the current sheet. 
However, this turns out not to be the case when the line is calculated as 
optically thin using data from the CHIANTI package \citep{2006ApJS..162..261L}.
Figure~\ref{fig:si_profile} shows that emission in the \FeXII\ line 
appears to consist of long strands outlining the canopy above the flux-emerging region,
showing mainly the long ($20$~Mm) strands or loops that lie above the shorter 
interacting loop systems described above in section~\ref{sec:resultEB}.
No additional \FeXII\ emission is visible in the vicinity of the 
current sheet. This is not because the current sheet does not emit in the 
\FeXII\ 19.51~nm line, but arises because absorption from the neutral 
hydrogen and helium gas overlies the current sheet, which has high 
opacity that is included in the calculation and is found to absorb the emission from this line.
Running the calculation without absorption does indeed show bright emission emanating
from the site of the current sheet. Recently, \citet{2019ApJ...871...82G} have found \FeXII\ emission in the (usually very weak) 134.9~nm line observed with IRIS, cospatial with \SiIV\ UV burst emission. This provides observational evidence for coronal temperatures in this event.

As shown in the central panel of figure~\ref{fig:si_profile}, the \SiIV\ 139.376~nm 
line has copious emission in the vicinity of the current sheet, up to two or more 
orders of magnitude brighter than the average intensity of this line in the 
computational domain. We find that the average line profile has a total intensity 
of 2~W/m$^2$/sr, an average redshift of 10~km/s, and a width of
25~km/s (which is reasonably close to thermal). The situation is found to be 
radically different in the area around the current sheet: As seen from above, the 
total intensity of this line reaches very high values in a thin region with 
a width of some hundred kilometers, sometimes extending to several hundred kilometers, 
stretching along the extent of the current sheet `leaf', roughly 
500--700~km. As noted and shown in figure~\ref{fig:ebuv_bz_tg_evol}, an 
extended region of cool gas lies above the current sheet. Because continuum absorption 
is weak, this cool gas is relatively transparent near 140~nm, but we find 
line absorption in the \SiIV\ 139.376~nm line when it becomes broad enough 
to be blended with cool 
lines such as the \NiII\ 139.332~nm line, 
which lies 93~km/s blueward of the \SiIV\ 139.376~nm line center. 
We show in figure~\ref{fig:si_profile} the line profile as calculated
with RH and including the opacity of \NiII. The \SiIV~line has 
attained high intensity and an extremely non-Gaussian shape with a width of
more than 200~km/s. Absorption in the \NiII~line is visible (the 
\SiIV~140.277~nm line profile has an almost identical shape, but shows no 
intensity dip at 93~km/s  blueward of the line center). This shows that enough cool 
material lies above the current sheet to produce the UV burst to both hinder 
emission in the  \FeXII\ 19.51~nm line and show absorption for a cool line such 
as \NiII\ 139.332~nm. The latter line is at times shifted 
5--10~km/s blueward, which is an indication that the 
bubble of cool gas carried by the emerging magnetic flux is still partially 
ascending at this time.

We consider the temperature and velocity structure of the upper portion of the 
current sheet in more detail. We find very high velocities throughout the 
current sheet at heights from 700~km to $3.5$~Mm. The upper part of the current 
sheet has upflow velocities of up to $200$~km/s in single-pixel columns, 
while the lower portion of the current sheet shows downflows of more than 
$-50$~km/s. This is clear from Figure~\ref{fig:ebuv_uztg_UV}, which shows 
$5\times 5$ grid-point averages of the vertical velocities and temperatures in and 
in the vicinity of the current sheet.
Reconnection in the current sheet 
drives bidirectional flows, and the highest temperatures are in locations 
where these flows are decelerated in shocks, near $z=1$~Mm for the downflows 
and $z=3$~Mm for the upflows. 
This plasma, heated to well above $100\,000$~K and moving in both 
directions at high velocity, gives rise to the extreme emission in transition 
region lines, such as shown above for the \SiIV\ lines around 140~nm. 
Figure~\ref{fig:emergence_side}, showing the total intensity of the 
\SiIV~$139.376$~nm line as seen from the side (and as calculated as optically thin 
using data from the CHIANTI atomic data package), reaches high values, up to two 
orders of magnitude above ambient, along the entire upper part of the current sheet 
from some 700~km to 3~Mm above the photosphere. 



\section{Discussion and conclusions}

The simulations presented in this paper show that the conditions required to 
generate many of the diagnostics associated with both EBs and UV bursts can occur 
naturally as a result of the emergence of a fairly strong but untwisted magnetic flux 
sheet into an atmosphere that contains a preexisting magnetic field of similar strength. The successfully synthesized diagnostics include \halpha\ 
wing enhancements in the form of `moustaches' and flame-like structures in \halpha\ 
spectroheliograms. We also found brightenings in the wings of the \CaII~triplet lines 
(as well as occasionally the line core), and the acceleration of cool 
surges was seen in the \CaII~triplet wings. 

We did not include phenomena 
that may be important to the physics or diagnostics of the chromosphere, 
such as ambipolar diffusion and nonequilibrium ionization. In addition, 
the magnetic diffusivity of the numerical model is, by necessity, several orders of magnitude higher than what is the case on the real Sun. 
However, the very good fit between the synthetic diagnostics and the 
observations and the nonexistent change in these diagnostics when the 
resolution was increased by nearly a factor 2 compared to the \citet{2017ApJ...839...22H} study, make us confident that this model is qualitatively correct.


Colocated and cotemporal with the EB discussed in this paper, 
we also found that the \SiIV\ lines near 140~nm were greatly enhanced, by two to three orders of
magnitude. They also showed extremely non-Gaussian line profiles and broadening, which is indicative 
of velocities of up to $>200$~km/s. The \FeXII\ 19.51~nm emission, on the other hand, shows 
no reaction to the presence of hot dense gas in the first few megameter above the photosphere.
The general picture that we obtain here is that when EBs and UV bursts
occur simultaneously as part of the same structure, EBs are formed in the first few 
hundred to one thousand kilometers of the upper photosphere and lower chromosphere, while the UV burst 
is formed up to several chromospheric scale heights higher in the chromosphere, over an extended 
region. In the specific example considered in this paper, the EB is formed from the photosphere up to 1200~km above, while the UV burst is formed at heights between 700~km and 3~Mm above the photosphere. The vertically elongated and flame-like shape of the EB is very similar to what is observed \citep[\ie{}][]{2011ApJ...736...71W}, while the observations of UV bursts are not as clear: \cite{2018SSRv..214..120Y} write that {\it ``\ldots a burst may appear with extended structure (jet, fibril, loop) connected to it, but these are typically less bright. The burst itself may also appear spatially extended into one direction `flame', but remains $<2$~arcsec''}. It remains to be determined whether this is compatible with the emission that this simulation produced. The observations indicate that between 10\% and 20\% of EBs show signatures of UV bursts \citep{2016A&A...593A..32G,2016ApJ...824...96T,Ortiz_etal2018}. We found some examples in the simulation described here, but not enough to carry out a statistical analysis. Furthermore, such an analysis 
should probably take into consideration a more realistic topology of the 
emerging field, for example, from an emerging active region- This is beyond the scope of this study. 

The source of this emission is the current sheet that is formed by the collision of two 
oppositely directed bundles of magnetic flux that are pushed together by photospheric motions. 
The current sheet stretches from the photosphere some 3 Mm into the 
chromosphere while spanning 1-2 Mm horizontally. It is the site of 
vigorous reconnection, accelerating bidirectional jets of plasma to velocities of 
about the Alfv\'en speed; tens of km/s near the photosphere up to several hundred km/s 
at heights of 1~Mm or more. While the \halpha\ and \CaII\ wing emission
is formed mainly in the first few hundred kilometers above the photosphere, we also find 
absorption by fast-moving gas at much greater heights,{} as seen in the blue wing of
\CaII\ in Figure~\ref{fig:ca_triplet_profile}, for example, where the cool surge is comprised of gas 
2~Mm above the photosphere. This is presumably gas accelerated by the Lorentz force, 
as first described by \citet{1996PASJ...48..353Y}, or with additional acceleration 
by pressure gradient forces as the upward-moving gas is pressed against a 
`wall' of strong field, as discussed by \citet{2016ApJ...822...18N}. The main
result of reconnection is the heating of plasma in the vicinity of the current sheet, 
either directly through Joule heating or resulting from the thermalization of the 
kinetic energy of the gas that is accelerated by reconnection jets as they are slowed down
in shocks. In the first few hundred kilometers above the photosphere, the gas is 
heated $2-3\,000$~K above ambient temperatures, but at greater heights, 
where radiative losses are less efficient, the plasma can at times attain temperatures of 
1~MK or more for several hundred seconds. 

The current sheet is located in a large bubble of emerging magnetic field, carrying 
with it cool gas from the photosphere. This cool gas has high 
opacity in typically photospheric or cool lines, but also in the continua of hydrogen 
and helium shortward of 91.1~nm, which is relevant for the EUV Imaging
Spectrometer for Hinode
and the Advanced Imaging Assembly on the Solar Dynamics Observatory observables such as 
\HeII\ 30.4~nm, \FeXII\ 19.51~nm, and \FeIX\ 17.1~nm. Thus we find emission from 
the current sheet suppressed in the \FeXII\ 19.51~nm line, and we find line 
absorption by \NiII\ 139.332~nm line in the wing of \SiIV\ 139.376 when this line 
becomes broad enough. 

During the course of the simulation, several regions of opposite magnetic field polarity 
were brought together as a result of photospheric motions. These led to 
reconnection in or just above the photosphere, in the middle to upper chromosphere, 
or in some cases in both when the topology of the interacting magnetic bubbles
allowed the formation of a current sheet with a length of several megameters. We therefore 
expect either EBs, UV bursts or both to be generated readily as a result of flux 
emergence. When both phenomena occur simultaneously or nearly so, as described in 
this paper, it may be possible to observe a certain shift in location of the EB 
and UV burst depending on the orientation of the current sheet and the viewing angle; such shifts would presumably be easier to confirm when they are observed close to the solar 
limb.

We therefore see no compelling reason to posit that UV bursts occur in the photosphere. Instead we propose that the observations point to a scenario where the chromosphere 
has become vastly bloated with slowly rising (10 km/s) cool fairly dense gas, up to 
10~Mm or more. Further, that this cool gas supplies the necessary opacity to explain absorption in lines such as \NiII\ and \MnI, and in the continua of hydrogen and neutral 
and singly-ionized helium. These provide narrow absorption bands in the \SiIV\ 
and \MgII\ line wings, and continuum absorption suppressing the evidence for 
intense heating in the AIA bands such as 30.3, 17.1, and 19.3~nm in the lower
regions of the expanded active region chromosphere. 


\begin{acknowledgements}
This research was supported by the Research Council of Norway through grant
170935/V30, through its Centres of Excellence scheme, project number 262622. 
Computing time has come through grants from the Norwegian Programme for Supercomputing, 
as well as from the Pleiades cluster through the computing project s1061, from the High End Computing (HEC) division of NASA.
The 3D radiative transfer 
computations were performed on resources provided by the Swedish National
Infrastructure for Computing (SNIC) at the High Performance Computing Center
North (HPC2N) at Ume\aa\ University and the PDC Centre for High Performance Computing
(PDC-HPC) at the Royal Institute of Technology in Stockholm.
Some images were 
produced by VAPOR ({\tt www.vapor.ucar.edu}), a product of the 
Computational Information Systems Laboratory at the National Center for Atmospheric Research. \end{acknowledgements}

%
\bibliographystyle{aa} 
\bibliography{solarrefs.bib} 
%

\end{document}